\documentclass[amsmath,floatfix,twocolumn,superscriptaddress]{revtex4-1}
\usepackage[utf8]{inputenc}
\usepackage{subfigure}
\usepackage{siunitx}
\usepackage{amssymb}
\usepackage{amsmath}
\usepackage{graphicx}
\usepackage{array}
\usepackage{dcolumn}
\usepackage{psfrag}
\usepackage{bm}
\usepackage{color}
\usepackage{multirow}

\begin{document}

\title{Machine Learning-Accelerated Band-Edge Engineering of Pnictogen Chalcohalide Solid Solutions for Solar Energy Technologies}

\author{Cibrán López}
    \affiliation{Departament de Física, Universitat Politècnica de Catalunya, 08034 Barcelona, Spain}
    \affiliation{Research Center in Multiscale Science and Engineering, Universitat Politècnica de Catalunya, Campus Diagonal-Besòs, Av. Eduard Maristany 10-14, 08019 Barcelona, Spain}
    \email{cibran.lopez@upc.edu}

\author{David Rovira}
    \affiliation{Barcelona Research Center in Multiscale Science and Engineering, Universitat Politècnica de Catalunya, 08019 Barcelona, Spain}
    \affiliation{Department of Electronic Engineering, Universitat Politècnica de Catalunya, 08034 Barcelona, Spain}

\author{Edgardo Saucedo}  
    \affiliation{Research Center in Multiscale Science and Engineering, Universitat Politècnica de Catalunya, Campus Diagonal-Besòs, Av. Eduard Maristany 10-14, 08019 Barcelona, Spain}
    \affiliation{Department of Electronic Engineering, Universitat Politècnica de Catalunya, 08034 Barcelona, Spain}

\author{Claudio Cazorla}
    \affiliation{Departament de Física, Universitat Politècnica de Catalunya, 08034 Barcelona, Spain}
    \affiliation{Research Center in Multiscale Science and Engineering, Universitat Politècnica de Catalunya, Campus Diagonal-Besòs, Av. Eduard Maristany 10-14, 08019 Barcelona, Spain}
    \affiliation{Institució Catalana de Recerca i Estudis Avançats (ICREA), Passeig Lluís Companys 23, 08010 Barcelona, Spain}
    \email{claudio.cazorla@upc.edu}

\begin{abstract}
Pnictogen chalcohalide (MChX; M=Bi,Sb; Ch=S,Se; X=I,Br) solid solutions combine earth-abundant constituents, tunable band gaps ($1.2$--$2.1$~eV), and strong optical absorption, making them attractive for solar energy conversion. Yet their vast compositional space has so far prevented a systematic assessment of how band-edge positions vary with stoichiometry and surface termination. Here, we combine first-principles density functional theory with machine learning to predict the valence and conduction band-edge positions of $\mathrm{Bi}_x\mathrm{Sb}_{1-x}\mathrm{S}_y\mathrm{Se}_{1-y}\mathrm{I}_z\mathrm{Br}_{1-z}$ solid solutions across their full compositional range on the two most stable surfaces, (010) and (011). We find that the valence-band maximum and conduction-band minimum can be tuned by more than $1$~eV through composition alone, and shift by up to $0.6$~eV between the two surface terminations for a same composition despite their nearly degenerate formation energies, establishing facet selection as a design parameter on par with chemical substitution. Guided by these results, we identify specific compositions capable of driving hydrogen, ammonia, methane, hydrogen peroxide, and oxygen (photo)electrochemical half-reactions, and show that several electron- and hole-transport contact materials commonly used in photovoltaic devices align with MChX solid solutions only as hole-selective contacts.
\end{abstract}

\maketitle

\section{Introduction}
\label{sec:intro}

Earth-abundant pnictogen chalcohalides (MChX; M=Bi,Sb; Ch=S,Se; X=I,Br) have recently garnered significant attention as a new family of semiconductors with remarkable intrinsic properties \cite{Ghorpade2023}. Their band gaps fall in the optimal range for solar energy conversion ($1.2$--$2.1$~eV), while their absorption coefficients ($25$--$66$~$\mu$m\textsuperscript{-1}) rival those of established photovoltaic absorbers \cite{Lopez2024,Ganose2016,Nielsen2025,Nayak2019}. They combine robust thermodynamic stability \cite{Nie2019,Li2024}, low-temperature processability (below $300^\circ$C) \cite{Cano2023,Li2024}, and non-toxicity, positioning them as promising candidates for a broad spectrum of energy-related applications.

\begin{figure*}[t]
  \centering
    \includegraphics[width=1.0\textwidth]{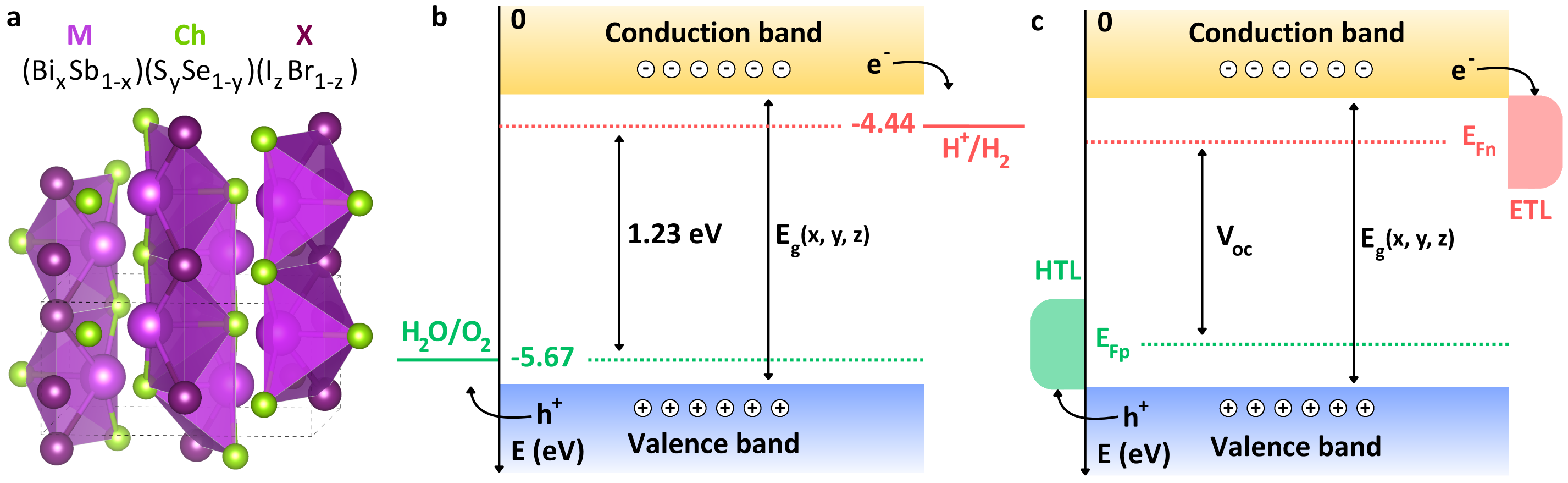}
    \caption{\textbf{Crystal structure and photoactivity representation of pnictogen chalcohalide solid solutions.} 
    \textbf{a.}~Ground-state orthorhombic phase with space group $Pnma$. 
    \textbf{b.}~Electronic band diagram suitable for photocatalytic water splitting. The conduction band minimum (CBM) and valence 
     band maximum (VBM) straddle the HER (H\textsuperscript{+}/H\textsubscript{2}) and OER (H\textsubscript{2}O/O\textsubscript{2}) 
     potentials, expressed with respect to the vacuum level, requiring a minimum band gap of $1.23$~eV. 
    \textbf{c.}~Electronic band diagram suitable for a photovoltaic device incorporating MChX as the light absorber layer and generic 
    selective hole (HTL) and electron transport (ETL) contact layers.}
    \label{fig1}
\end{figure*}

While ternary MChX compounds (Fig.~\ref{fig1}a) already exhibit very promising optoelectronic properties, solid solutions with general formula Bi$_{x}$Sb$_{1-x}$S$_{y}$Se$_{1-y}$I$_{z}$Br$_{1-z}$ open new avenues for property engineering, such as continuous tuning of band gaps and dielectric responses \cite{Lopez2024}. Such compositional tunability is particularly valuable for optimizing device performance in various critical energy applications such as photovoltaics (e.g., tandem solar cells) and photocatalysis. 

In photocatalysis, the ability to shift band edges relative to redox potentials determines the feasibility of driving chemical transformations like hydrogen and oxygen evolution reactions \cite{Guo2019,Xu2000} (Fig.~\ref{fig1}b). In photovoltaics, band edge engineering facilitates optimal absorber-selective contact alignments, maximizing sunlight harvesting and charge extraction \cite{Colombara2020} (Fig.~\ref{fig1}c). Solid solutions therefore may provide a powerful chemical space for engineering pnictogen chalcohalides beyond the capabilities of their ternary counterparts.

The full landscape of MChX solid solutions contains an enormous number of possible atomic configurations, each potentially giving rise to distinct surface structures and electronic properties. Exhaustive exploration using first-principles density functional theory (DFT) is computationally prohibitive, while experimental synthesis and characterization of all possible compositions are equally impractical. This represents a general challenge in functional materials discovery, where high-throughput screening and data-driven approaches are increasingly employed to navigate vast compositional spaces \cite{Jain2013,Kiyohara2024}. 

To address this challenge, in this study we leverage machine learning (ML) to accelerate the exploration of MChX solid solution surfaces with first-principles accuracy. By generating a large-scale dataset consisting of DFT simulations, we train predictive ML models that capture subtle structure-property relationships across this materials family. This systematic ML-aided investigation of band alignments reveals that compositional substitution enables significant and controlled tuning of band edge positions. As representative case studies, we explore the impact of MChX solid-solution band-edge positioning on photovoltaic devices and photocatalytic activity for green hydrogen and other fuel production, providing specific compositions for each of these applications. Beyond advancing the fundamental understanding of MChX solid solutions, our work establishes a generalizable ML-accelerated framework for probing surfaces in emergent semiconductors, bridging the gap between atomistic modeling and device-level design.

\begin{figure*}[t]
  \centering
    \includegraphics[width=\textwidth]{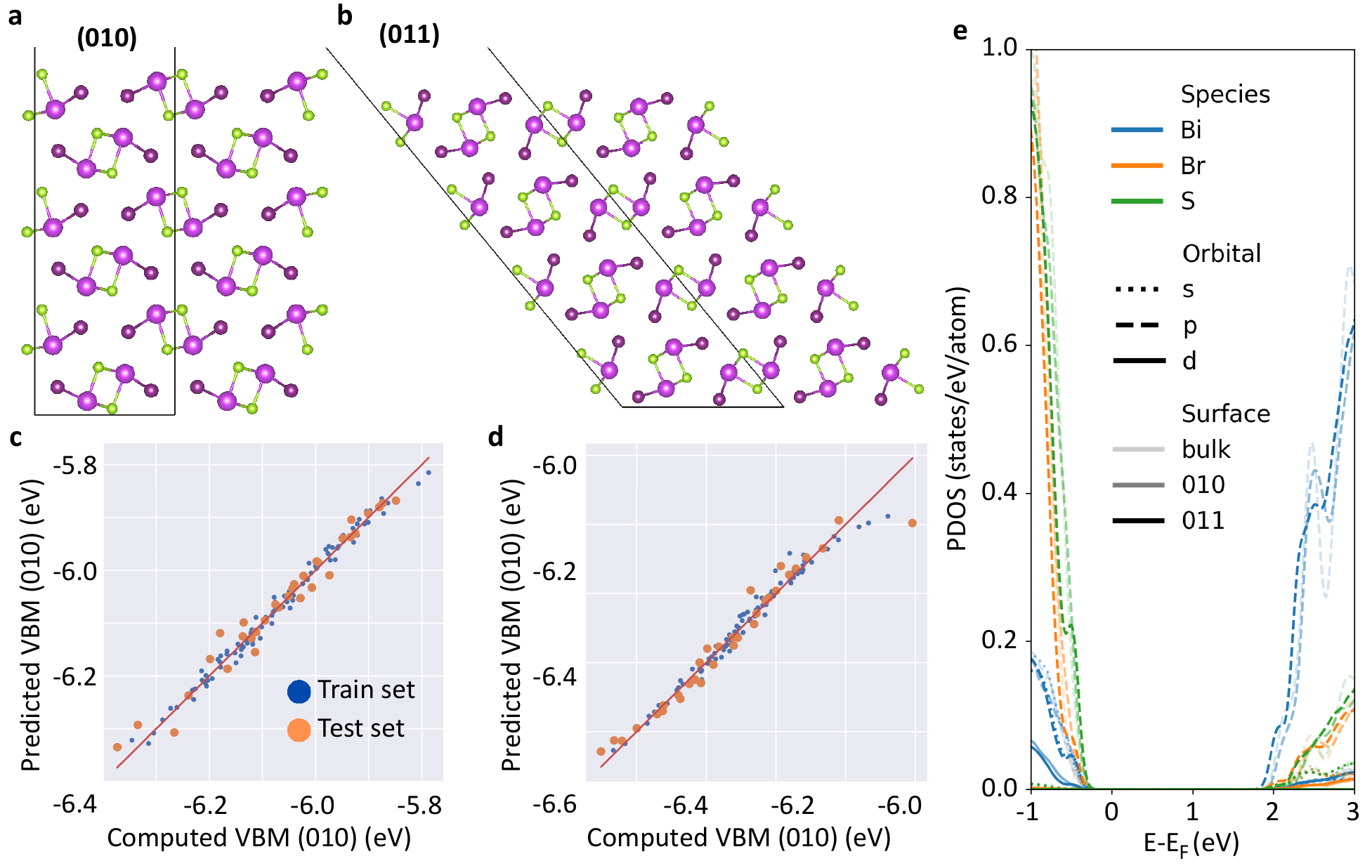}
    \caption{\textbf{MChX surface terminations and ML training and validation test.}
    \textbf{a,b.}~The (011) and (010) surfaces are the most favorable terminations, with nearly identical formation energies. 
    \textbf{c,d.}~Parity plots for the developed ML models predicting the VBM with respect to vacuum level considering the (011) 
    and (010) surfaces.}
    \label{fig2}
\end{figure*}

\section{Results and Discussion}
\label{sec:results}

MChX semiconductors crystallize in an orthorhombic phase (space group $Pnma$) consisting of quasi-one-dimensional columns weakly coupled by van der Waals interactions (Fig.~\ref{fig1}a) \cite{Cano2025,Nielsen2025,Lopez2025,Lopez2025-3}; in the present study, this structure is assumed to be the stable phase of MChX solid solutions as well. Both theoretical calculations and experimental data confirm that MChX compounds are thermodynamically stable against decomposition into competing secondary phases \cite{Cano2023,Li2024}. First-principles DFT calculations for MChX (Methods) also yield excellent agreement with experiments across their structural, vibrational, and electronic properties \cite{Cano2025}. Within this theoretical framework, the (010) (Fig.~\ref{fig2}a) and (011) (Fig.~\ref{fig2}b) surfaces are found to be the most energetically favorable in ternary MChX \cite{Lynch2026}; in the present study, they are assumed to remain the lowest-energy surfaces in MChX solid solutions as well.

In the following sections, we present theoretical results for the band alignments of Bi$_{x}$Sb$_{1-x}$S$_{y}$Se$_{1-y}$I$_{z}$Br$_{1-z}$ solid solutions with arbitrary composition $\{x,y,z\}$. These results are then analyzed in the context of potential photovoltaic and photocatalytic applications: in photovoltaics, band alignment dictates whether charge carriers are extracted or lost to recombination \cite{Colombara2020}; in photocatalysis, it determines whether a surface can straddle the redox potentials required for water splitting or CO$_{2}$ reduction \cite{Xu2000,Guo2019}. 

Our results were obtained using a predictive ML model trained on an extensive DFT dataset of valence band maximum (VBM) levels calculated with respect to vacuum. Results from analogous predictive ML models for the band gap and formation energy of MChX solid solutions developed in a previous work \cite{Lopez2024} were used to (i)~derive the conduction band minimum (CBM) levels with respect to vacuum and (ii)~select those systems that are thermodynamically stable against phase decomposition. ML predictions were carried out on a dense compositional grid with intervals of $0.05$ in $x$, $y$, and $z$, encompassing a total of $9,261$ solid solutions.

\subsection{Dataset generation and ML models performance}
\label{subsec:dataset-ml}

An extensive database of first-principles DFT calculations \cite{Cazorla2017} covering Bi$_{x}$Sb$_{1-x}$S$_{y}$Se$_{1-y}$I$_{z}$Br$_{1-z}$ solid solutions with compositions $\{x,y,z\} = \{0, 0.25, 0.5, 0.75, 1\}$, totaling $125$ compounds, was constructed. Geometry relaxations for the two representative (010) and (011) surfaces were performed by employing the Perdew-Burke-Ernzerhof exchange-correlation energy functional revised for solids (PBEsol) \cite{Perdew2008} (Methods). To estimate band alignments, the range-separated hybrid functional HSEsol \cite{Heyd2003,Schimka2011} was employed along with spin-orbit coupling (SOC) corrections \cite{Lopez2024,Nielsen2025,Lopez2025,Lopez2025-3}.

Modeling of solid solutions and alloys poses technical challenges due to the requirement of multiple and large simulation supercells to accurately replicate arbitrary stoichiometries and chemical disorder \cite{Shenoy2019,Tuli2023}. Simulation techniques like the special quasirandom structure (SQS) approach \cite{Zunger1990} have significantly simplified the modeling of chemically disordered crystals by introducing radial correlations in site occupations. However, the size of SQS supercells can still be prohibitively large for the HSEsol+SOC calculations pursued in this study. For this reason, here we resort to the virtual crystal approximation (VCA, Methods), a computationally efficient approach that has been proved successful for modeling isoelectronic substitutions in MChX solid solutions and semiconducting materials in general \cite{Lopez2024,Eckhardt2014}.

We trained an ML multilayer perceptron (MLP) model on our database of $125$ pnictogen chalcohalides to accurately predict the surface band alignments of MChX solid solutions with respect to vacuum (Figs.~\ref{fig2}c,d). The performance of the trained ML-MLP model is fairly good for the two analyzed surfaces. In particular, valence band predictions display cross-validation train and test mean absolute errors of 12.3 and 8.9 meV, respectively, for the (011) surface, and 7.3 and 4.8 meV, respectively, for the (010) surface. These results allow predicting the full landscape of solid solutions with high fidelity.

\begin{figure*}[t]
  \centering
    \includegraphics[width=1.0\textwidth]{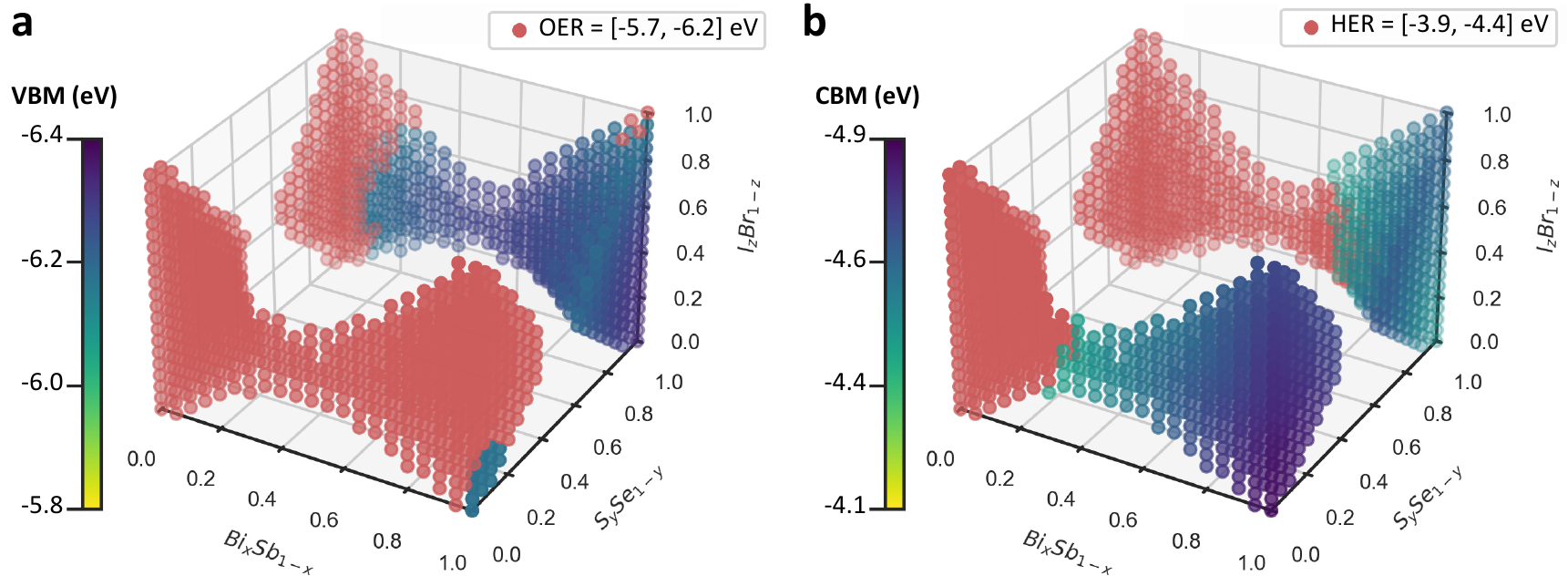}
    \caption{\textbf{Band alignment results obtained for MChX solid-solutions considering the (010) surface.}
    \textbf{a.}~VBM and \textbf{b.}~CBM predicted values. Results are only shown for compositions that render thermodynamically stable
    systems against decomposition into secondary phases \cite{Lopez2024}. Predicted VBM and CBM values that are suitable for OER and HER 
    catalysis, respectively, are highlighted.}
    \label{fig3}
\end{figure*}

\begin{figure*}[t]
  \centering
    \includegraphics[width=1.0\textwidth]{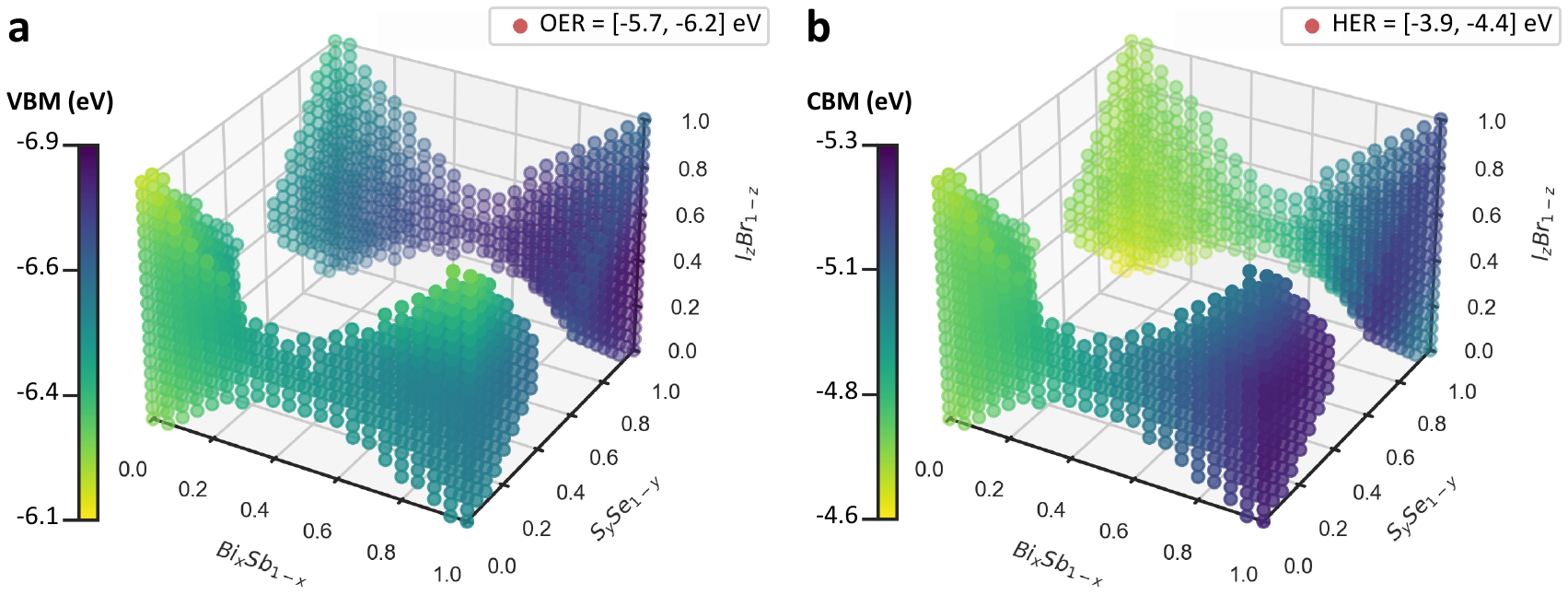}
    \caption{\textbf{Band alignment results obtained for MChX solid-solutions considering the (011) surface.}
    \textbf{a.}~VBM and \textbf{b.}~CBM predicted values. Results are only shown for compositions that render thermodynamically stable
    systems against decomposition into secondary phases \cite{Lopez2024}. Predicted VBM and CBM values that are suitable for OER and HER 
    catalysis, respectively, are highlighted.}
    \label{fig4}
\end{figure*}

\subsection{Band alignments}
\label{subsec:band}

Our ML-aided first-principles calculations reveal a continuous spectrum of band alignments for synthesizable MChX solid solutions. In this study, we considered only compositions previously predicted to lie on, or very close to, the convex hull \cite{Lopez2024}. For the (010) surface (Fig.~\ref{fig3}), the VBM spans from -6.33~eV for BiSI\textsubscript{0.25}Br\textsubscript{0.75} to -5.80~eV for SbSeI, while the CBM ranges from -4.87~eV for BiSeBr to -4.17~eV for SbSI\textsubscript{0.15}Br\textsubscript{0.85}. For the (011) surface (Fig.~\ref{fig4}), the corresponding ranges are -6.88~eV for BiSI\textsubscript{0.25}Br\textsubscript{0.75} to -6.17~eV for SbSeI for the VBM and -5.25~eV for BiS\textsubscript{0.15}Se\textsubscript{0.85}I\textsubscript{0.5}Br\textsubscript{0.5} to -4.62~eV for SbSBr for the CBM. Interestingly, although the (010) and (011) surfaces of the parent compounds are nearly degenerate in energy (i.e., $\Delta \gamma_{\rm surf} = \gamma_{\rm (010)} - \gamma_{\rm (011)} = 0.012-0.024$~J/m$^{2}$, depending on the material), they can exhibit markedly different band alignments. For example, a maximum difference of $0.58$~eV between the two surfaces is predicted for BiSI\textsubscript{0.4}Br\textsubscript{0.6}. Meanwhile, Bi\textsubscript{0.95}Sb\textsubscript{0.05}SeI displays minimum VBM and CBM differences for the (010) and (011) surfaces of 0.39 eV.

The origin of such large variations in band alignment across different surfaces may involve multiple factors. Surface-dependent effects such as atomic relaxation, surface reconstruction, changes in local coordination environments, and the formation of surface dipoles can all significantly modify the electrostatic potential at the surface, thereby shifting the positions of the VBM and CBM relative to the vacuum level \cite{sharma22}. This argument is supported by the species- and orbital-resolved density of states shown in Fig.~\ref{fig2}e, computed for bulk, (010), and (011) environments: the valence band edge is dominated by S and Br $p$ states with a smaller Bi $s$+$p$ (lone-pair) contribution, while the conduction band edge is dominated by Bi $p$ states, and this orbital composition is essentially unchanged across all three environments when referenced to each system's own Fermi level. These electronic features indicate that neither surface introduces new electronic states or qualitatively alters the bonding character at the band edges; instead, the pronounced shifts reported above must originate from a rigid displacement of the electrostatic reference level itself, consistent with a surface-dipole origin.

As discussed earlier, the optoelectronic properties of MChX semiconductors make them attractive candidates for both photovoltaics \cite{Cano2025,Lopez2025,Lopez2025-3}, where efficient light harvesting and charge-carrier extraction are essential for high device performance, and photocatalysis, where sunlight is used to drive chemical reactions such as water splitting \cite{Ghorpade2023,Nielsen2025}. In both applications, the positions of the VBM and CBM relative to the vacuum level are critical parameters for achieving optimal performance. In the following sections, we present and discuss the results of our computational investigations of MChX solid solutions in the context of these two important energy-conversion technologies.

\begin{figure*}[t]
  \centering
    \includegraphics[width=1.0\textwidth]{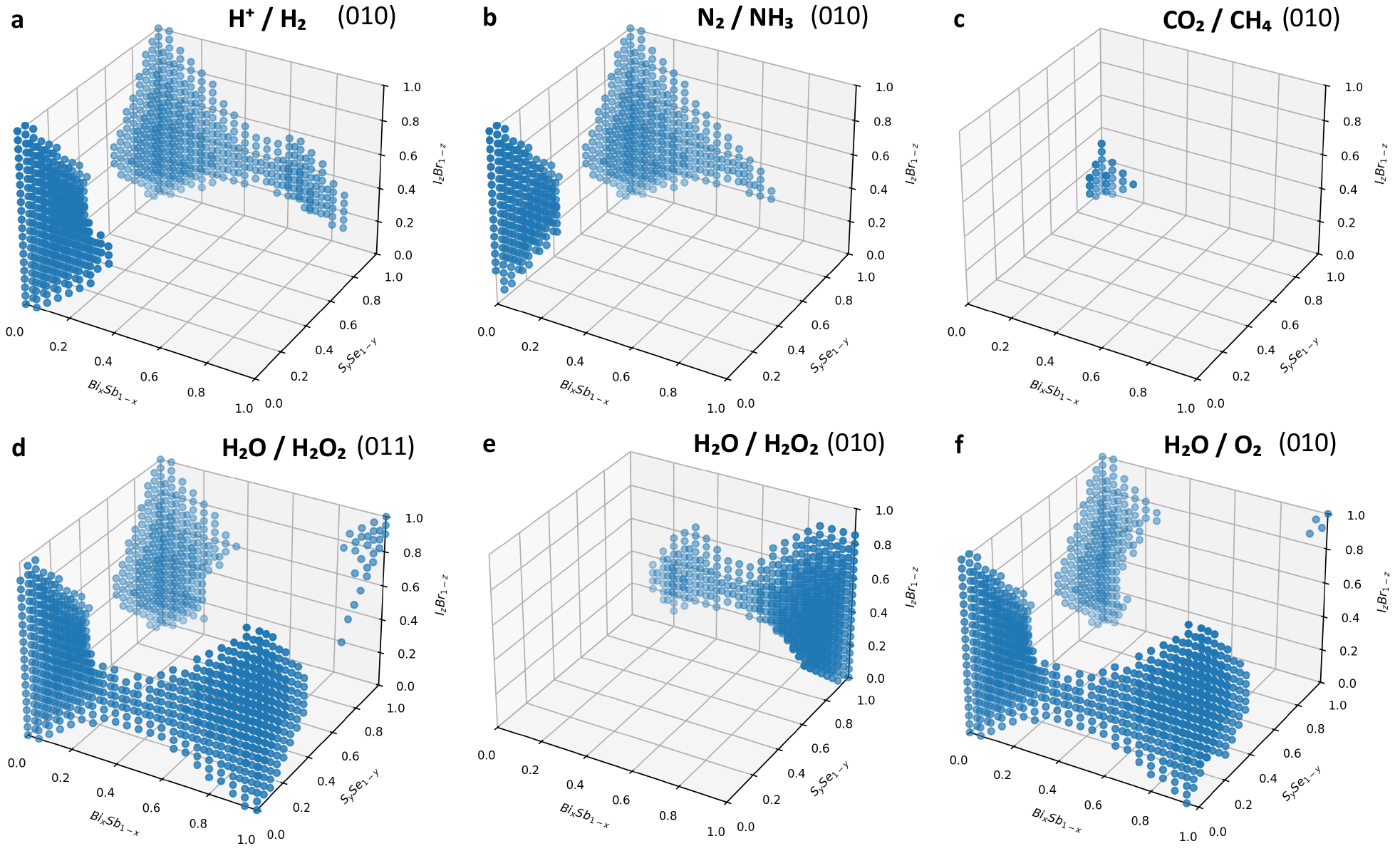}
    \caption{\textbf{Fuel-related chemical reactions, including water splitting, for which (010) and (011)-oriented MChX solid-solutions 
    may be promising photocatalysts.}
    Results are only shown for compositions that render thermodynamically stable systems against decomposition into secondary phases 
    \cite{Lopez2024}.}
    \label{fig5}
\end{figure*}

\subsection{Application~1: Photocatalysis}
\label{subsec:photocatal}

For photocatalytic applications, molecular adsorption must be energetically favorable, ideally with a free energy close to zero so that it can proceed spontaneously as part of the reaction pathway (the Sabatier principle \cite{sabatier}). In addition, the conduction and valence band edges must straddle the relevant redox potentials so that photogenerated electrons and holes possess sufficient thermodynamic driving force to reduce and oxidize the adsorbed species, respectively. For water splitting, for example, this principle requires the CBM to lie above the hydrogen evolution potential ($-4.4$~eV with respect to vacuum) and the VBM to lie below the oxygen evolution potential ($-5.7$~eV with respect to vacuum) \cite{wateredox}. Since band edge positions are surface properties sensitive to atomic termination and crystallographic orientation, even small surface-dependent shifts can decide whether a given orientation is catalytically active or inactive.

Figure~\ref{fig3} shows the band-alignment results obtained for MChX solid-solutions considering the (010) surface orientation. In general, VBM and CBM values remain almost unchanged when varying halogen content, with the VBM and CBM shifting only slightly across the full $\mathrm{I}_z\mathrm{Br}_{1-z}$ range, with variations which amount up to 0.02~eV (Bi$_{0.15}$Sb$_{0.75}$SI$_z$Br$_{1-z}$, $z$: 0.15 $\to$ 0.2) and 0.03~eV (Bi$_{0.65}$Sb$_{0.35}$SI$_z$Br$_{1-z}$, $z$: 0.45 $\to$ 0.5) for the VBM and CBM, respectively. The VBM is instead most sensitive to the pnictogen ratio becoming markedly deeper as bismuth content increases at the expense of antimony, with differences of up to 0.29~eV (Bi$_x$Sb$_{1-x}$S$_{0.15}$Se$_{0.75}$I$_{0.1}$Br$_{0.9}$, $x$: 0 $\to$ 1), with a comparable but slightly weaker deepening produced by increasing sulfur content at the expense of selenium, with differences of up to 0.27~eV (Bi$_{0.15}$Sb$_{0.75}$S$_y$Se$_{1-y}$I$_{0.1}$Br$_{0.9}$, $y$: 0 $\to$ 1). The CBM, on the other hand, is governed almost entirely by the pnictogen ratio, becoming substantially deeper with increasing bismuth content by up to 0.50~eV (Bi$_x$Sb$_{1-x}$S$_{0.15}$Se$_{0.75}$I$_{0.1}$Br$_{0.9}$, $x$: 0 $\to$ 1), while increasing sulfur content instead makes the CBM shallower, partially offsetting the bismuth-driven deepening by up to 0.35~eV (Bi$_{0.9}$Sb$_{0.1}$S$_y$Se$_{1-y}$I$_{0.05}$Br$_{0.95}$, $y$: 0 $\to$ 1). The clearest departure from this general behaviour occurs at selenium-rich compositions, where the antimony end shows a somewhat shallower VBM than expected, and the bismuth end shows a somewhat deeper CBM than expected, suggestive of a synergistic bismuth-selenium interaction.

The suitability of (010)-oriented MChX solid solutions for light-driven water splitting is further illustrated in Fig.~\ref{fig3}. (In the following analysis, a tolerance of $0.5$~eV above and below the HER and OER potentials, respectively, is adopted to account for overpotential requirements). On the (010) surface, a wide range of compositions is suitable for OER (Fig.~3a): this includes selenium-rich compositions ($y \approx 0$ in $\mathrm{S}_y\mathrm{Se}_{1-y}$), compositions along the $\mathrm{SbSI}_x\mathrm{Br}_{1-x}$ solid solution, and a small number of compositions near $\mathrm{BiSI}$. HER (Fig.~3b), by contrast, is favorable mainly for antimony-rich stoichiometries ($x \approx 0$ in $\mathrm{Bi}_x\mathrm{Sb}_{1-x}$), with bismuth incorporation additionally tolerated when sulfur is the dominant chalcogen ($y \approx 1$ in $\mathrm{S}_y\mathrm{Se}_{1-y}$).

Figure~\ref{fig4} shows the band-alignment results obtained for MChX solid-solutions considering the (011) surface orientation. As on the (010) surface, halogen substitution has the weakest influence on both quantities, leaving the VBM and CBM nearly unchanged across the full iodine-bromine range by as much as 0.03~eV (Bi$_{0.9}$Sb$_{0.1}$S$_{0.05}$Se$_{0.95}$I$_z$Br$_{1-z}$, $z$: 0.8 $\to$ 0.85 for the VBM and Bi$_{0.75}$Sb$_{0.25}$SeI$_z$Br$_{1-z}$, $z$: 0.55 $\to$ 0.6 for the CBM, respectively). The VBM responds most strongly to the chalcogen ratio, deepening as sulfur replaces selenium by up to 0.36~eV (Bi$_{0.95}$Sb$_{0.05}$S$_y$Se$_{1-y}$I, $y$: 0 $\to$ 1), with a somewhat weaker deepening from increasing bismuth content of up to 0.25~eV (Bi$_x$Sb$_{1-x}$S$_{0.8}$Se$_{0.2}$I$_{0.7}$Br$_{0.3}$, $x$: 0 $\to$ 1), while the CBM remains dominated by the pnictogen ratio, deepening substantially with increasing bismuth content by up to 0.48~eV (Bi$_x$Sb$_{1-x}$S$_{0.65}$Se$_{0.35}$I$_{0.45}$Br$_{0.55}$, $x$: 0 $\to$ 1) and only partially offset by increasing sulfur content, by up to 0.20~eV (Bi$_{0.9}$Sb$_{0.1}$S$_y$Se$_{1-y}$I$_{0.05}$Br$_{0.95}$, $y$: 0 $\to$ 1). The clearest exception to this general trend occurs at the bismuth-rich, selenium-rich, iodine-rich corner of the composition space, where both the VBM and the CBM are noticeably shallower than the rest of the dataset would suggest, indicating that this particular combination of substitutions partly counteracts the strong band-deepening effect otherwise associated with bismuth incorporation.

As shown in Fig.~\ref{fig4}, (011)-oriented MChX solid solutions are not suitable photocatalysts for water splitting, as they globally fail to straddle the HER and OER potentials. This behaviour stands in stark contrast to that observed for (010)-oriented MChX solid solutions (Fig.~\ref{fig3}), and the underlying reasons for such substantial differences in band alignment were already discussed in Sec.~\ref{subsec:band}. Rather than representing a limitation, this marked photocatalytic contrast between (010) and (011) surfaces highlights the unique tunability and rich compositional versatility offered by MChX solid solutions.

Figure~\ref{fig5} summarises fuel-related chemical reactions for which (010)- and (011)-oriented MChX solid solutions may serve as promising photocatalysts; for completeness, water splitting has been included as well. Three reduction half-reactions are considered on the (010) surface (Figs.~5a--c): proton reduction, $2\mathrm{H^{+}} + 2e^{-} \rightarrow \mathrm{H_2}$ ($\Delta E = E - E_{\mathrm{vac}} = -4.44$~eV \cite{Trasatti1986}); nitrogen reduction to ammonia, $\mathrm{N_2} + 6\mathrm{H^{+}} + 6e^{-} \rightarrow 2\mathrm{NH_3}$ ($\Delta E = -4.35$~eV \cite{Huang2021}); and carbon dioxide reduction to methane, $\mathrm{CO_2} + 8\mathrm{H^{+}} + 8e^{-} \rightarrow \mathrm{CH_4} + 2\mathrm{H_2O}$ ($\Delta E = -4.20$~eV \cite{Habisreutinger2013}).

Suitable compositions for both H$_2$ evolution (Fig.~5a) and NH$_3$ synthesis (Fig.~5b) are dominated by antimony-rich stoichiometries ($x \approx 0$ in $\mathrm{Bi}_x\mathrm{Sb}_{1-x}$), with bismuth incorporation additionally tolerated when sulfur is the dominant chalcogen ($y \approx 1$ in $\mathrm{S}_y\mathrm{Se}_{1-y}$) for the (010) surface. This window narrows sharply for CH$_4$ (Fig.~5c), where only compositions near $\mathrm{SbSBr}$ satisfy the shallower conduction-band requirement, reflecting how demanding this eight-electron reduction is relative to H$_2$ or NH$_3$ production for the (010) surface.

Two oxidative pathways for water are considered in parallel, each requiring a valence-band maximum deeper than its corresponding potential: the two-hole route to H$_2$O$_2$, $2\mathrm{H_2O} \rightarrow \mathrm{H_2O_2} + 2\mathrm{H^{+}} + 2e^{-}$ ($\Delta E = -6.20$~eV \cite{Tang2024}), evaluated on both the (011) and (010) surfaces, and the four-hole route to O$_2$, $2\mathrm{H_2O} \rightarrow \mathrm{O_2} + 4\mathrm{H^{+}} + 4e^{-}$ ($\Delta E = -5.67$~eV \cite{Trasatti1986}). The H$_2$O$_2$ window on the (011) surface (Fig.~5d) is accessible for selenium-rich compositions ($y \approx 0$ in $\mathrm{S}_y\mathrm{Se}_{1-y}$) as well as compositions along the $\mathrm{SbSI}_x\mathrm{Br}_{1-x}$ solid solution, together with a small number of compositions near $\mathrm{BiSI}$. On the (010) surface (Fig.~5e), the accessible compositions shift instead to the $\mathrm{BiSI}_x\mathrm{Br}_{1-x}$ solid solution, indicating that the harder two-hole oxidation pathway favors bismuth-rich, sulfur/iodine-dominant stoichiometries on this particular facet. Molecular oxygen evolution on the (010) surface (Fig.~5f) reproduces the same set of regions found for H$_2$O$_2$ on the (011) surface: selenium-rich compositions, the $\mathrm{SbSI}_x\mathrm{Br}_{1-x}$ solid solution, and a small number of $\mathrm{BiSI}$-like compositions.

\begin{figure*}[t]
  \centering
    \includegraphics[width=1.0\textwidth]{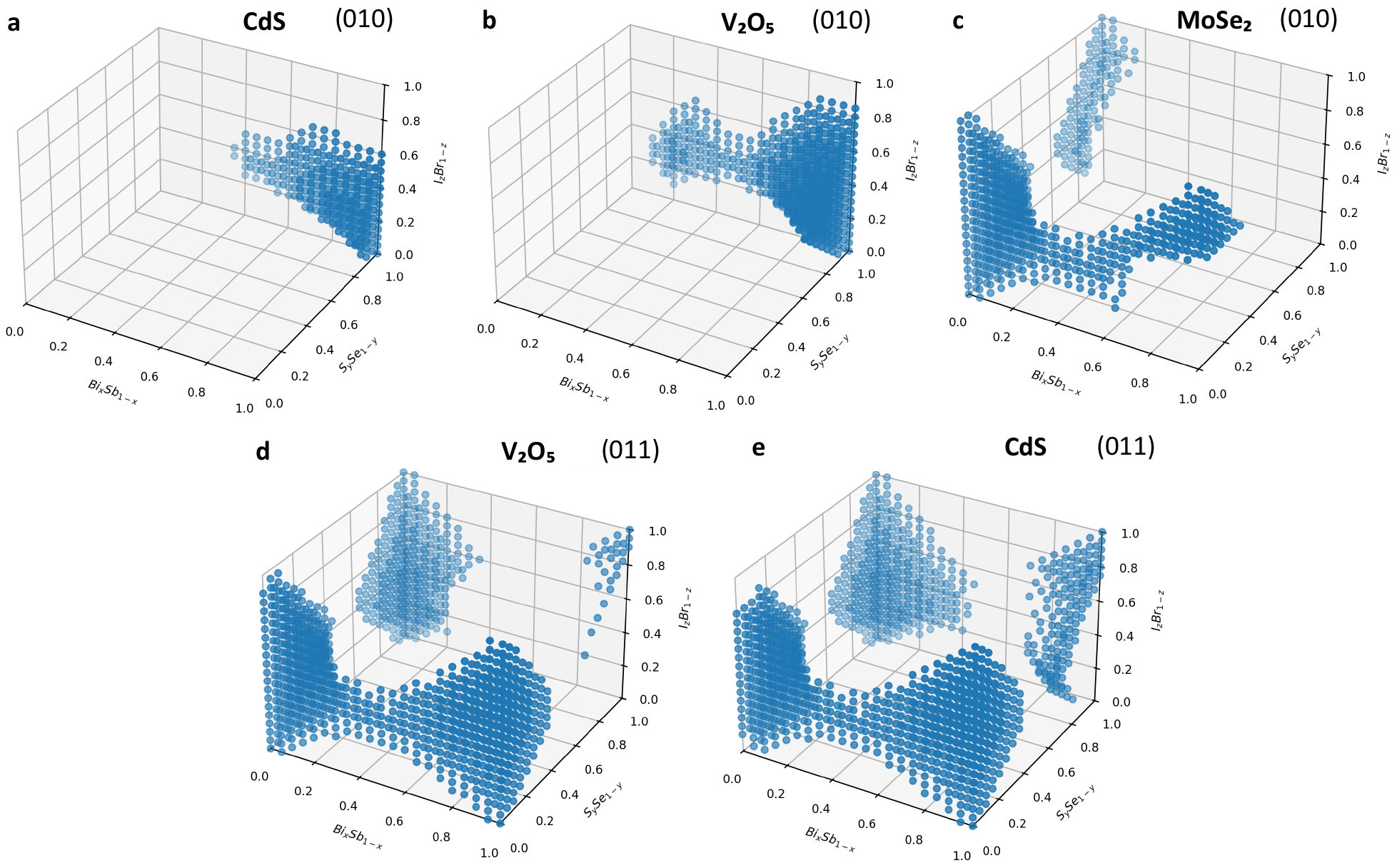}
    \caption{\textbf{Charge-extraction compatibility of several typical selective-contact materials with MChX solid-solution light absorbers.}
    Results are only shown for compositions that render thermodynamically stable systems against decomposition into secondary phases
    \cite{Lopez2024}. It is found that CdS, V$_2$O$_5$ and MoSe$_2$ can act exclusively as HTL for MChX solid solutions.}
    \label{fig6}
\end{figure*}

\subsection{Application~2: Photovoltaics}
\label{subsec:photovoltaics}

Although the intrinsic optoelectronic properties of MChX suggest excellent potential for photovoltaic performance, experimental power conversion efficiencies remain below $10$\%, well short of their Shockley-Queisser limit of $\sim 30$\% \cite{Cano2025,Nielsen2025}. While non-radiative recombination at point defects likely accounts for part of this discrepancy \cite{Lopez2025,Lopez2025-3}, device-level limitations associated with surface-dependent band alignments may be equally important \cite{Klein2020}. In particular, precise knowledge of band alignment at device interfaces is crucial to ensure that selective contact layers extract photogenerated carriers efficiently (Fig.~\ref{fig1}c); unfavorable alignments promote spontaneous recombination before those carriers can contribute to photocurrent.

Fig.~\ref{fig6} shows the potential compatibility in terms of charge extraction of several typical selective contact materials (i.e., CdS, V$_2$O$_5$ and MoSe$_2$) with MChX solid-solution light absorbers. In particular, a material is considered a suitable hole-transport layer (HTL) if its reported VMB lies between the absorber VBM and 0.5~eV below it, whereas it is considered a suitable electron-trasnport layer (ETL) if its reported CMB lies between the absorber CBM and 0.5~eV above it.

Regarding (010)-oriented MChX solid solutions, the range of potentially compatible HTL materials is broad, as all three reference materials show a viable HTL alignment. CdS ($\Delta E_{\mathrm{VBM}} = E_{\mathrm{VBM}} - E_{\mathrm{vac}} = -6.27$~eV \cite{Dimitriev2018}, Fig.~\ref{fig6}a) is compatible with bismuth-rich, sulfur-rich, bromine-rich compositions ($x \approx 1$, $y \approx 1$, $z \approx 0$), and V$_2$O$_5$ ($\Delta E_{\mathrm{VBM}} = -6.20$~eV \cite{Meyer2011}, Fig.~\ref{fig6}b) reproduces this same region with a slightly wider compositional margin. MoSe$_2$ ($\Delta E_{\mathrm{VBM}} = -5.60$~eV \cite{Shimada1994}, Fig.~\ref{fig6}c), in contrast, is compatible with selenium-rich compositions ($y \approx 0$) and performs markedly worse toward the bismuth-rich end, with only a small number of additional matches along the $\mathrm{SbSI}_x\mathrm{Br}_{1-x}$ solid solution. These three HTL candidates are therefore complementary rather than redundant: MoSe$_2$ covers the antimony- and selenium-rich portion of the compositional space, while CdS and V$_2$O$_5$ cover the bismuth-, sulfur-, and bromine-rich end, together spanning nearly the full (010) compositional range, giving considerable flexibility in selecting a hole-selective contact matched to a specific MChX alloy composition.

For (011)-oriented systems, the range of compatible contacts is comparatively more limited: only two of the three reference contact materials show a compatible band alignment, both as HTLs. V$_2$O$_5$ (Fig.~\ref{fig6}d) is compatible with a compositional window resembling that found for MoSe$_2$ on the (010) surface, together with a small number of compositions near $\mathrm{BiSI}$, while CdS (Fig.~\ref{fig6}e) reproduces the same set of regions but over a still wider compositional range.

Across both surface orientations, however, no ETL match is found for any of the three reference materials considered here, so (011)-oriented devices are constrained to hole-extraction engineering at both faces of the absorber, and even the broader HTL flexibility available on the (010) surface does not resolve this gap. Achieving a functional architecture will therefore require identifying a genuinely electron-selective contact from outside this set, or engineering the conduction-band offset of one of these materials (e.g.\ via doping or interfacial dipole modification) to enable electron extraction, for either orientation.

\section{Conclusions}
\label{sec:conclusions}
The integration of MChX into advanced energy conversion devices presents promising opportunities due to their high thermodynamic stability, favorable optoelectronic properties and low-temperature synthesis. However, challenges persist, including the vast unexplored landscape of MChX compositions and the disconnect between understanding optoelectronic properties and designing energy conversion devices. To address these challenges, we have proposed a solution combining ML techniques and first-principles DFT calculations, to efficiently predict the surface-dependent band-alignment properties of MChX solid solutions so as to optimize the design of photocatalytic and solar-cell devices. 

In particular, our ML-accelerated first-principles framework shows that the VBM and CBM can be tuned by several tenths of an electronvolt through chemical substitution alone, and shift by a further, comparable amount between the (010) and (011) surface terminations while conserving the underlying band gap, establishing facet selection as a design parameter on par with composition itself. This tunability translates into sharp application-specific outcomes: the (010) surface supports unassisted water splitting, with antimony-rich compositions favoring H$_2$/NH$_3$ evolution and sulfur-/bismuth-rich compositions favoring O$_2$/H$_2$O$_2$ generation, whereas the (011) surface fails to straddle the HER/OER potentials despite forming at a nearly identical surface energy. For photovoltaics, all three tested contacts (CdS, V$_2$O$_5$, MoSe$_2$) align only as hole-transport layers, together covering nearly the full (010) compositional range but leaving electron-selective contact design an open problem. Taken together, these results establish MChX solid solutions as a compositionally and facet-tunable platform for targeting at least five distinct fuel-forming photocatalytic reactions and hole-selective photovoltaic contacts within a single materials family.

The findings presented in this study not only advance the fundamental understanding of MChX solid solutions but also validate an innovative and universally applicable approach for the design of solar cells and photocatalytic devices, showcasing the potential of MChX materials in advancing energy conversion technologies. This ML-accelerated framework is transferable to other emergent 
semiconductor families, opening broader avenues for surface-property-guided materials design.

\section*{Methods}
\textbf{First-principles calculations.}~\textit{Ab initio} simulations based on density functional theory (DFT) were performed to analyze the physicochemical properties of bulk pnictogen chalcohalides. These calculations were conducted with the \verb!VASP! software package \cite{Kresse1996} using the generalized gradient approximation to the exchange-correlation energy for solids due to Perdew \textit{et al.} (PBEsol) \cite{Perdew2008}. Spin-orbit coupling effects, which are particularly relevant for Bi-based MChX \cite{Ganose2018,Ganose2016,Lopez2023,Lopez2024-1}, were taken into account for the calculation of optoelectronic properties along with range-separated hybrid functionals containing an exact Hartree-Fock exchange fraction of $25$\% (i.e., HSEsol+SOC \cite{Schimka2011,Heyd2003,Krukau2006}). The projector augmented-wave method was used to represent the ionic cores \cite{Blochl1994} and the following electronic states were considered in valence: Bi $6p, 5d, 6s$; Sb $5p, 4d, 5s$; Se $4p, 4s$; S $3p, 3s$; I $5p, 5s$; Br $4p, 4s$. Wave functions were represented in a plane-wave basis truncated at $450$~eV. By using these parameters and dense $\Gamma$-centered \textbf{k}-point grids for reciprocal space Brillouin zone integration, the resulting energies were converged to within $1$~meV per formula unit. In the geometry relaxations, a tolerance of $0.5$~meV$\cdot$\AA$^{-1}$ was imposed in the atomic forces. 
\\

\textbf{Surfaces generation.}~For the generation of surfaces and input files for a given input bulk structure, we used our in-house developed \verb!SurfAIcing! software \cite{SurfAIcing}. By relying on foundational ML-IAPs, \verb!SurfAIcing! is able to relax surfaces and automatically postprocess all the required DFT data to generate surface formation energy rankings and band-alignments. 
\\

\textbf{Solid-solution modeling.}~We employed the virtual crystal approximation (VCA) \cite{Bellaiche2000}, which assumes the presence of virtual atoms on potentially disordered sites, interpolating between the electronic properties of the actual components. VCA supercells match the size of the primitive cell, making them computationally manageable, and have been shown to provide reliable optoelectronic results for isoelectronic atom substitutions \cite{Eckhardt2014,Zhu2020,Prandini2018}. This approximation is particularly well-suited to the MChX family because the substituted atoms (Bi/Sb, S/Se, I/Br) are isoelectronic and share similar bonding character, minimizing local structural disorder effects that would otherwise invalidate the VCA.
\\

\textbf{Band-alignment calculations.}~To estimate accurate band alignments, we followed the first-principles computational strategy employed in previous works \cite{Liu2020,Liu2022}. Briefly, both bulk and slab calculations were performed from which the alignment of the electrostatic potential within the dielectric material can be obtained relative to the vacuum level. From the slab calculations, the average electrostatic potential of the material relative to the vacuum level was obtained. From the bulk calculations, the valence band maximum and conduction band minimum relative to the average electrostatic potential were determined. These calculations involved the estimation of macroscopic and planar average potentials. The planar potential was computed by averaging potential values within a well-defined plane (for instance, perpendicular to the surface of the slab), and the macroscopic potential was obtained by taking averages of the planar potential over distances of one unit cell along the chosen direction. The slab systems should be thick enough to ensure that the electron density in the center of the slab is practically equal to that in the bulk material. We found that $\approx 15$~\AA~ thick slabs accompanied by similarly large portions of vacuum provided sufficiently well converged electrostatic potentials.
\\

\textbf{Machine learning models.}~Multilayer perceptrons as implemented in the Scikit-learn package \cite{Pedregosa2011} were used for the prediction of the energy and optoelectronic properties of MChX solid solutions \cite{Kiyohara2024,Lopez2025-2,Benitez2025-4}. Specifically, a multilayer perceptron regressor with two layers of 32 and 64 neurons each was implemented. The input consisted of the relative abundance of each element in the solid solution, and the output was the targeted property. The input data was normalized using a standard scaler. Regarding the hyperparameters, the best performance was achieved with an Adam optimizer with a constant learning rate of 0.001, a rectified linear unit (ReLU) activation function, and $\alpha = 0.05$ strength for the L2 regularization term of the loss function. The performance of the models was evaluated through the mean absolute error (MAE) obtained from a k-fold cross-validation strategy applied on both the training and test datasets (with k=20, as this is common practice in the literature and provides statistically significant results). ML-aided first-principles DFT predictions were made on a dense grid of Bi$_{x}$Sb$_{1-x}$S$_{y}$Se$_{1-y}$I$_{z}$Br$_{1-z}$ compositions taken at $0.05$ intervals on $x$, $y$ and $z$. This comprehensive approach enabled us to assess the surface energetics and band alignments of a total of $9,261$ solid solutions. A repository containing our ML-MLP models and python codes has been created and is freely accessible at \cite{codes}.
\\

\section*{Data availability}
The data that support the findings of this study have been made publicly available \cite{database}, comprising the MChX band alignments of all simulated reference 125 solid-solutions.
\\

\section*{Acknowledgments}
C.L. acknowledges support from the Spanish Ministry of Science, Innovation and Universities under a FPU grant. C.C. acknowledges support by MICIN/AEI/10.13039/501100011033 and ERDF/EU under the grants CNS2025-165467, PID2023-146623NB-I00 and PID2023-147469NB-C21 and by the Generalitat de Catalunya under the grants 2021SGR-00343, 2021SGR-01519 and 2021SGR-01411. Computational support was provided by the Red Española de Supercomputación under the grants FI-2024-1-0005, FI-2024-2-0003, FI-2024-3-0004, FI-2024-1-0025, FI-2024-2-0006, and FI-2025-1-0015. This work is part of the Maria de Maeztu Units of Excellence Programme CEX2023-001300-M funded by MCIN/AEI (10.13039/501100011033). E.S. acknowledges the European Union H2020 Framework Program SENSATE project: Low dimensional semiconductors for optically tunable solar harvesters (grant agreement Number 866018), Renew-PV European COST action (CA21148) and the Spanish Ministry of Science and Innovation ACT-FAST (PCI2023-145971-2). E.S. is grateful to the ICREA Academia program. 
\\

\section*{Author contributions}
C.L. and C.C. conceived the study and planned the research, which was discussed in-depth with the rest of the co-authors. C.L. performed the first-principles calculations and analyzed the results. The manuscript was written by C.L. and C.C., with substantial input from the rest of the co-authors.
\\

\section*{Competing financial interests}
The authors declare no competing financial interests.
\\

\bibliography{Bibliography}

\end{document}